# Numerical study of the normal current density behaviour in a narrow -gap glow discharge

S.A. Starostin[1,✉], P.J.M. Peters[2,✉], E.Kindel[1], A.V. Azarov[2], S.V. Mitko[2], and K.J.Boller[2]

[1] *Istitut für Niedertemperatur Plasmaphysik, Friedrich-Ludwig-Jahn-Str. 19*
*17489 Greifswald, Germany*

[2] *University of Twente, Laser and Non-linear Optics Group,*
*Faculty of Technical Sciences, PO Box 217, 7500 AE, Enschede, The Netherlands*

**Abstract**

*A numerical study of normal glow discharge properties was performed in the case of small electrodes separations (0.05 – 0.4 cm) and moderate gas pressures (10 - 46 Torr). A recently observed new experimental effect of a considerable reduction in the normal current density for smaller discharge lengths was analyzed both by means of 2D fluid model and by a minimal 1D drift model of gas discharge. A good agreement between theoretical and experimental behaviour was demonstrated. An influence of the electrodes separation and of the gas heating on the value of the normal current density is discussed.*

✉ **E-mail:** starostine@inp-greifswald.de, p.j.m.peters@utwente.nl

**Introduction**

The normal glow discharge investigation history spans more than a century thus one could expect that this is one of the best-studied subjects in modern physics. Still, in a variety of conditions and applications where normal discharges are used (e.g. lighting, plasma processing and laser pumping) the physics of the normal glow discharge remains an intensively discussed issue.

According to the generally adopted view [1,2], a diffusive, self-sustained discharge between two parallel plane electrodes can be classified as follows: at very small currents (a few μA or less) when the space charge does not affect the electric field distribution, the discharge occupies the whole electrode surface. Such discharge is called a Townsend discharge. The sustaining voltage is equal to the breakdown voltage and does not change with the current. As the current gradually increases the influence of space charge becomes more pronounced and the discharge begins to contract. The slope of the voltage-current characteristic in this regime is negative (causing instabilities). This discharge is called subnormal. When the current density reaches a certain value, with a further current increase the discharge begins to spread again. While the total current increases, the current density and voltage remain constant. This is known as the "normal" discharge and is characterised by the so-called normal current density, the normal voltage drop and the normal cathode layer thickness. As soon as the current spot on the cathode will cover the whole surface of the electrode, the voltage will start to rise again with a further increase in current. This is known as the "abnormal" regime. For small values of the *pd* parameter (*pd* < $pd_c$, were $pd_c$ is usually in order of 1 Torr.cm) a normal phase does not exist and a direct transition from Townsend to abnormal discharge, with increasing current, occurs. In this case the discharge will always occupy the whole electrode surface.

The development of micro-discharges for UV light sources, plasma display elements or as active medium for gas lasers recently led to a renewed interest in discharges occurring in short gaps at elevated gas pressures. It appears that there is a lack of information on glow discharges in discharge gaps with the *pd* value in the order of $pd_c$ and with the discharge in a transient state between "subnormal" and "normal". Moreover in relation to the present interest for bifurcation phenomena in gas discharges, the relaxation to a "normal glow" at low *pd* values can be considered as an important mechanism for structured discharge formation [3]. The process of normal cathode spot appearance is important not only for continuous wave (direct current) operation, but for pulsed or alternating current (AC) and barrier discharges as well. The transition from Townsend to normal discharge was already studied in a number of theoretical and experimental articles [4,5,6,7,8]. However these papers are not focused on the properties of the normal discharges itself.

The most important scaling parameter characterizing a normal discharge is the reduced normal current density $J_n/p^2$, where $J_n$ is the normal current density and $p$ the pressure. It is believed (see for instance [1]) that the value of $J_n/p^2$ is solely dependent on both the gas and electrode material and is independent of the discharge length. It should be noted that the theoretical verification is based on the classical, one-dimensional theory of the cathode layer developed in the first half of $20^{th}$ century by von Engel and Steenbeck [9]. However, as we found in recent experiments [10], $J_n/p^2$ becomes dependent on the discharge length, when the distance between the electrodes is in order of the negative glow length.

The classical discharge model was further developed by Ward in the late 50s - early 60s [11,12]. Ward performed one of the first numerical investigations of the gas discharge, making self-consistent calculations of the electric field and drift current. However this work was by far complete and no special investigation was done to study the behaviour of the $J_n/p^2$ value versus the $pd$ parameter. The recent comprehensive analytical and numerical analysis of the 1-dimensional discharge model done in [3] reveals new types of voltage-current characteristics at low $pd$ values. However, the normal current behaviour was not studied in this work.

It is important to note that within the 1-dimensional model the normal current density can not be calculated directly. It is generally accepted to follow the assumption, made by von Engel and Steenbeck, that the normal current corresponds to the abscissa of the minimum on the calculated voltage-current characteristic. However this assumption remains a constantly argued issue (see [1]) and is not proven yet. Strictly speaking the formation of the normal current spot is at least a two-dimensional effect so its description requires adequate models.

One of the earliest two-dimensional numerical study of the normal discharge in nitrogen was done by Raizer and Surzhikov in [13]. At that time 2D calculations were very time consuming but even this somewhat incomplete two-dimensional modelling contributed a lot to a deeper insight on the process of the normal current spot formation [1].

More recent numerical investigations of the normal discharges and of the transition from Townsend to normal regimes for a $pd$ range of 1-10 Torr cm were presented in [4,5]. The numerical study done in [4] with a 2D fluid model both with a local and non-local ionization term was mainly focused on the low current Townsend and subnormal regimes. In the same article a comprehensive semi-analytical model of the current spot was proposed. Particularly it was predicted that existence region of the normal current spot can be extended to lower *pd* values if secondary emission is a growing function of the electric field [4]. In [5] the applied numerical model of the discharge is similar to the one described in the present work, but without accounting for gas heating. It will be shown below that gas heating can significantly influence the properties of normal discharge especially at moderate gas pressures. The modelling in [5] was performed for a discharge in Ar at a fixed inter-electrode distance of 1 cm and for two different values of the gas pressure (3 and 6 Torr). By increasing the discharge current a transition from Townsend to subnormal, to normal and to an abnormal discharge was followed. Additionally subnormal oscillations were investigated. Nevertheless from the results presented in [4,5] it is difficult to make a conclusion about the behaviour of the normal current density in the vicinity of the value $pd_c$, therefore an additional investigation is needed.

Despite the impressive development of gas discharge models, especially during last years, such a fundamental parameter as the normal current density was not studied in detail. So, the knowledge on the normal current density behaviour and its scaling properties hardly changed since the classical discharge model was proposed by von Engel and Steenbeck seventy years ago. This situation is also caused by the lack of new experimental information on the subject.

In the present work a two-dimensional fluid model of the discharge was applied to analyse the recent experimental results [10]. At the same time we made calculations using the classical discharge model [11] in order to see if a minimal approach to discharge modelling can reflect the newly observed behaviour of the normal current density versus *pd*. It should be additionally noticed that although in this work we are considering the same range of *pd* parameter as in [4,5], here it corresponds to the case of small electrode separations and moderate gas pressures. These conditions are interesting for many nowadays applications of glow discharge.

**Modelling**

The primary goal of the model developed here was to investigate if the experimentally observed decrease of the normal current density with the gap length at short electrode distances [10] can be understood by means of one of the standard approaches that have so far been used to simulate gas discharges. It is known that discharges in rare gases are influenced in their qualitative behaviour by impurities. The experiments described in [10] were carried out without gas flow thus a contamination of the gases used in these experiments is not excluded. That is why we decided to model a discharge in air for the sake of quantitative comparison with experimental data. Additionally, we made calculations of the normal current densities at low *pd* values, utilizing Ward's 1D minimal model of the gas discharge [11].

A fully spatial dependent simulation of the discharge geometry with rectangular-shaped electrodes used in [10] would require a 3 dimensional model in Cartesian co-ordinates. However, as it follows from the measurements [10], the values of the normal current density are independent on the electrode width. At the same time it is important to show by means of modelling that the dependency of the normal current density on the discharge length is not exclusively a property of the discharge geometry applied in [10]. From this consideration for our modelling we have chosen a more general 2D cylindrical geometry, keeping gas pressure and electrode separation corresponding to [10]

According to the numerical results of kinetic simulations [4] the assumption of a local dependency of the ionisation coefficient on the electric field is a good approximation for subnormal discharges, while for normal discharges at *pd* ranges of 1-10 Torr.cm the non-local nature of the ionisation coefficient becomes more pronounced. In our modelling we are using a drift-diffusion approximation where the ionisation coefficient, electron mobility and electron diffusion coefficient are functions of the mean electron energy. This approach still allows accounting for "small" non-local effects. The possible influence of run-away electrons on the discharge characteristics remains out of consideration.

A 3-moment drift-diffusion fluid equation set (see Appendix A and [14]) was applied for modelling of the normal discharge in air in a two-dimensional cylindrical geometry. The corresponding set of equations includes the continuity equations for electrons and positive ions, the equation for the mean electron energy, the Poisson equation for the electric potential, and the gas thermal conductivity equation.

Though in general the electron energy distribution function (EEDF) hardly resembles a Maxwellian one, according to ref. [14], it is possible to introduce the mean electron energy as the main characteristic of the distribution function and to consider all of the coefficients as being dependent only on this mean energy. To derive the mean electron energy, a corresponding equation that takes into account the energy loss, the gain and the electron heat conduction has to be solved (A4). The rate and transport coefficients as functions of the mean electron energy are determined beforehand by numerical solving the 0D Boltzmann equation for the EEDF. Then these dependencies, in the form of tabulated functions, are used for resolving equation set A1-A12. This approach requires the assumption that these rates and coefficients depend on the mean electron energy in the same way as they do in the equilibrium condition, when the local electron energy losses are balanced by the local energy gain [14].

For a discharge in air as well as for discharges in atomic gases, at different pressures a similar behaviour of the normal current density versus electrode separation was observed [10]. On the basis of these experimental observations it can be assumed that this behaviour is not directly related to volumetric processes as recombination or electron attachment. Therefore in our model we are limiting ourselves to a simple source term in the continuity equations including only one ionization process. Certainly, an in depth numerical study of a discharge in air would require accounting also for negative ions and for various plasma-chemical reactions but this is not the aim of the present work.

According to our calculations in the studied range of temperatures, the temperature dependency of the thermal conductivity coefficient $\lambda$ can be approximated by a linear function. (See Appendix A7).

Firstly we calculated the tabulated dependency of the thermal conductivity coefficient on the gas temperature and then deduced the coefficients A and B for the linear function (A7). Further, it was assumed that there is no convective gas flow in the discharge volume and that the local density can be determined from the ideal gas equation.

Our numerical method is based on an exponential discretisation scheme [15] of the drift-diffusion equations on a non-equidistant mesh. The transport and Poisson equations were integrated successively in time. Discretisation of the Poisson and continuity equations on a two dimensional grid, results in 5-points linear systems that were solved with the SIP (strongly implicit procedure) method [16]. Starting from initial conditions the evolution of the discharge parameters were advanced in time till a stationary solution was reached. A semi-implicit representation of the source terms in the Poisson and electron energy equation [17] was used to allow larger time steps.

The discharge was modelled in a cylindrical numerical domain with a radius of 0.7 cm. (The lateral size of this domain is not important as long as the cathode spot is smaller than the domain). The distance between the electrodes was chosen in the range of 0.5 to 4 mm.

Various non-equidistant meshes were used to ensure that the grid size does not affect the value of calculated normal current density (like it was observed in [13]).

The kinetic coefficients for electrons were calculated with the freeware BOLSIG program [18]. For the discharge in air we assumed a gas mixture of $N_2 : O_2 : Ar = 78 : 21 : 1$; the gas pressure in our calculations was varied from 10 to 46 Torr. The mobility of the positive ions was taken as 1.5 cm$^2$/(V sec) which corresponds to a gas density of $2.69 \; 10^{19}$ cm$^{-3}$. A value of 0.1 was taken for the secondary ion–electron emission coefficient $\gamma$. The dependency of the gas thermal conductivity coefficient $\lambda$ [W/cm K] on the temperature (see Appendix A7) was approximated with $A_{air} = 5.5 \; 10^{-7}$, $B_{air} = 1.18 \; 10^{-4}$.

It should be noticed that the dependency of the effective secondary electron emission coefficient $\gamma$ on the electric field value at the cathode is now an intensively discussed issue [5,19,20] and can significantly influence the results of the fluid model [5,17]. According to a recent comprehensive analysis [19] various mechanisms such as ion or electron emission, photo-emission by resonance photons, emission caused by meta-stables and fast atoms contribute to the total effective electron emission from the cathode. The influence of each process is dependent on the value of the *E/N* parameter on the cathode as well as on the state ("clean" or "dirty") of the cathode surface. If one considers only the process of secondary ion-electron emission then $\gamma$ is a fast growing function of the *E/N* parameter at low *E/N* values but it reaches a constant value at high *E/N*. The nature of this functionality has its origin in the effect of electron back-scattering towards the cathode. This dependency can considerably affect the sustaining voltage of the Townsend discharge and the properties of the subnormal discharge. It also shifts the minimal *pd* at which the normal discharge can exist to the lower values. Yet because, the $\gamma$ dependency on the electric field is unknown for a discharge in air, we have decided to use, for a first modelling, a constant $\gamma$ and thus do not take into account back-scattering effects. A study of the influence of the variation of $\gamma$ with the electric field will be a subject for further investigation.

The equation set for the 1-dimensional classical drift model is listed in Appendix B. Taking into account the continuity equation for the drift current and the electrostatic equation for the electric field results in two first order differential equations. The numerical solution of system (Appendix B: B1-B3) is performed by a "shooting" method where an initially chosen electrical field value at the cathode (or anode) is varied until the boundary conditions (B3) are met. For the calculation of the normal current density the generally accepted assumption that normal current corresponds to the minimum on the calculated voltage current characteristic was used.

The input parameters for the 1D drift model are as follows: the mobility of positive ions and the secondary electron emission coefficient were taken the same as for the 2D fluid model. The mobility of the electrons was chosen as 434 cm$^2$/(V sec) at a gas density of 2.69 10$^{19}$ cm$^{-3}$ (according to the results of the BOLSIG code [18] for a $N_2$ - $O_2$ - Ar gas mixture and 4 eV electron energy). Two cases were considered for the Townsend's ionization coefficient. In the first case the ionization coefficient was calculated as a function of the reduced electric field $E/p$ and then used as a tabulated function. In the second case according Raizer's textbook [1] an exponential approximation was used (see Appendix B4; with $s = 1$, $C = 15$, $D = 365$ for air ).

The current density profiles calculated with the 2D fluid model at the cathode for a discharge length of 1 mm are presented in Fig. 1 for three different values of the total discharge current $I$ = 7, 12 and 17 mA. One can see that a characteristic normal cathode spot is formed with a uniform current density in the centre and rather steep edges. Also can be seen that an increase of the discharge current results in a larger cathode spot, while the current density (and voltage drop) remains the same.

The experimental [10] and different theoretical dependencies of the reduced normal current density $J_n/p^2$ versus $pd$ parameter are presented in Fig. 2. One can see that the reduced normal current density decreases for smaller discharge gaps in all theoretical curves. The calculation is done for different gas pressures and the results are presented in this figure for 10, 20 and 46 Torr. In order to illustrate the influence of gas heating some calculations were done without the effect of gas heating, for a constant gas temperature of 290 K (curves 4 and 5). The results obtained using the 1D drift model are plotted for two different ionisation coefficient approximations. An analysis of these results can be found in discussion section.

The contour maps of the equi-potential lines for a normal discharge in air at 20 Torr and for electrode separations of 0.5, 1, 2 and 4 mm calculated with the 2D model are presented in Fig.3 (a-d). It can be seen that with the reduction of the inter-electrode gap, while the normal current density decreases (see also Fig.2, curve 2) the length of the cathode fall region increases.

In Fig.4 (a-d) the calculated contour plots for constant ionization rates (source term in the continuity equation A1: $k_i(\varepsilon)n_e N$ [cm$^{-3}$ / sec]) are shown for different discharge lengths, corresponding to potential distribution at Fig.3 (a-d). It can be seen that for the smallest electrode separation of 0.5 mm ($pd$ = 1 Torr cm) when the normal current spot still exist the cathode fall length (Fig.3(a)) as well as the high ionization region (Fig.4(a)) become of the same order as discharge gap. It should be noted though that already for an inter-electrode distance of 1 mm ($pd$ = 2 Torr cm) both the cathode voltage fall length and length of the region where the main ionization occurs are significantly smaller than total discharge length, yet the current reduction is still apparent for this electrode separation (see Fig.2, curve 2). Therefore the existence of the normal discharge is limited by a $pd_c$ value which is in order of cathode fall dimension $pd_n$. At the same time the dependency of the normal current density on the electrode separation can be seen for a range of $pd$ values which significantly exceeds $pd_n$.

**Discussion**

In [10] it was observed that the normal current density depends on the distance between the electrodes in the $pd$ range of a few Torr.cm: the $J_n$ decreases with reduction of the discharge length. For increasing discharge length the normal current density attains a constant value. It seems that this dependency on discharge length is of general nature because it was observed for different gases, pressures and electrode materials.

So far it was commonly assumed that the normal current density does not depend on the discharge length. This assumption was mainly based on a classical, semi-analytical model of the normal cathode layer proposed by von Engel and Steenbeck [9] (some aspects of this theory are discussed in [1]). In their model the field distribution inside the discharge was approximated by a linear function. The absolute value of the electric field linearly reduces from $E = E_0$ at the cathode surface ($x = 0$) to $E = 0$ at $x = d_s$, where $d_s$ is the sheath thickness. For $x > d_s$ the electric field is assumed to be zero. The ionisation coefficient was dependent on the local value of the electric field. From such approximation immediately follows that the sheath properties are independent on the discharge length if the distance between the electrodes is larger than the sheath thickness.

In the drift model developed by Ward [11] no 'a priori' given electric field profile was implied. Nevertheless the calculated field inside the cathode fall region was well approximated by a linear function which was in agreement with the experimental observations and confirming the previous assumption made by von Engel and Steenbeck [9]. At the same time the resulting electric field was at no place across the discharge gap zero, because this is a strict requirement for the drift current continuity. As it was mentioned in the introduction Ward's model for a long time remained insufficiently analysed. Particularly this is valid for the dependency of the $J_n/p^2$ parameter on the discharge length. Our calculations show that the current density, which corresponds to the minimum on the voltage-current curve (attributed to the normal regime), shifts toward lower values for shorter discharge lengths (see curves (6) and (7) in Fig 5). This dependency coincides with the one found in our experiment. It should be noted here that a one dimensional model can not directly describe the normal current spot. However, even such a simple discharge simulation approach gives already a valuable indication about the functionality between $J_n/p^2$ value and the inter-electrode distance.

In comparison with the classical one dimensional approach the 2-dimensional fluid model applied here is accounting for the drift-diffusion transport of the charged particles in a self-consistent 2D electrical field and also is characterised by the non-local ionization and transport coefficients which depend on the mean electron energy value.

A two-dimensional analysis is essential for the description of the radial behaviour of the cathode spot. According to [4] the shape of current spot is governed by the ion drift in the radial electric field for normal discharges and the radial electron diffusion for subnormal discharges. The last regime is characterized by considerably lower current densities in comparison with the normal discharge. The reduction in electrode separation for a normal discharge leads to a decrease in the radial electric field component at the current spot boundary (see Fig.3). The shape of current spot for a shorter discharge length is to a greater extent governed by electron diffusion which is a fast process. If the voltage current characteristic of such short discharge still has a minimum, then a normal regime may exist, but it will have a lower current density.

Two arguments should be considered when analyzing the results of the present numerical modelling concerning the normal current density behaviour. Firstly, already the simple 1D drift model gives a dependency of the $J_n/p^2$ parameter on the discharge length which is a consequence of the self-consistent solution for the transverse field and drift current. Secondly, for different electrode separations the balance conditions at the current spot boundary are changing, resulting, in turn, to a change in the sustained current density.

The influence of gas heating on the normal current density can be seen by comparing curves (2) and (4) in Fig 5, calculated respectively with and without taking into account the effect of gas heating. Gas heating leads to a decrease of the $J/p^2$ parameter. The reduction becomes significant at higher currents (and consequently higher heating rates).

This effect can be understood by the fact that the current density $J$ scales not as $J/p^2$ but, more precisely, is proportional to the square value of the gas concentration ($N^2$). With a constant gas pressure a higher gas temperature results in a lower gas density and therefore lower current density. The same reason is valid for the deviation from the $J/p^2$ scaling for the curves calculated for different gas pressures (see curves (2) and (3) in Fig.2). When gas heating is not taken into account the $J/p^2$ scaling holds (see curves (4) and (5) in Fig.2).

It can be also seen from Fig.2 that the calculated curves (2) and (3) are reproducing the experimental behaviour (curve (1)) very well, although having a lower absolute value of the $J/p^2$ parameter. This difference between curves (1) and (3) is approximately of factor of 2. This can be attributed to the simplicity of the applied model which did not take into account attachment and recombination processes. It should be also noted that the calculations and experiments were done for different discharge geometries. It can be expected, according to Fig.2, that different regimes of heat removal can cause a considerable variation in absolute values of current density.

It can be argued that fast non-local electrons created in the cathode fall are producing a significant amount of ion-electron pairs in the negative glow which can extend considerably beyond the cathode fall region. In this case the reduction in discharge length will diminish the electron multiplication while the electrode separation still greatly exceeds the size of cathode voltage fall. A detailed study of the influence of fast electrons on the formation of the discharge with a certain value of the normal current density would require a pure kinetic or hybrid model (see for instance [21]). This will be a subject for further investigation. However the non-local fluid model applied in the present work already gives a sufficiently good description of the experimental data.

It is important to note that the fact, that the normal current density depends on the distance between the electrodes, imposes certain requirements on the experimental methodologies and set-ups for the measurement of "handbook" values of the $J_n/p^2$ parameter. Therefore the results of previous measurements need to be analysed from this point of view.

**Summary**

The normal glow discharge was investigated numerically in a narrow gap geometry for a *pd* range of 1 - 10 Torr.cm at moderate gas pressures. A considerable reduction in current density for small discharge lengths was recently experimentally observed in [10] for different gases and electrode materials. It was found that such fundamental discharge similarity parameter as reduced normal current density $J_n/p^2$ becomes a function of the electrode separation at low *pd* values. While it was traditionally assumed that $J_n/p^2$ is defined only by gas and electrode material. In present work this new effect was analyzed with a 3-moments 2D fluid model. The value of $J_n/p^2$ was found to be strongly dependent on discharge length in the case when gas heating is neglected. An influence of the gas heating on the $J_n/p^2$ value and scaling properties were discussed. A good qualitative agreement was shown between modelling and experimental results. It was also shown that already a minimal 1D drift discharge model can give an indication of the $J_n/p^2$ functionality on the discharge length. We believe that more insight on the effect can be gained by analysing the influence of the non-local ionization with a hybrid (kinetic-fluid) modelling which can be a subject for further investigation.

# APPENDIX A

## Equation set for the two-dimensional fluid model of a DC discharge

**1 Continuity equations for positive ions and electrons:**

$$\frac{\partial n_{p,e}}{\partial t} + \text{div}(F_{p,e}) = k_i(\varepsilon) n_e N \tag{A1}$$

$$F_p = n_p \mu_p E - D_p \text{grad}(n_p) \tag{A2}$$

$$F_e = -n_e \mu_e(\varepsilon) E - D_e(\varepsilon) \text{grad}(n_e) \tag{A3}$$

In these equations is $n_{p,e}$ the density of positive ions and electrons respectively, $k_i(\varepsilon)$ the ionization coefficient, $F_{p,e}$ the charged particle flux, $N$ the neutral gas density, $\mu_{p,e}$ is the mobility, $D_{p,e}$ the diffusion coefficients and $E$ the electric field.

**2 Equation for the mean electron energy $\varepsilon$:**

$$\frac{\partial n_e \varepsilon}{\partial t} + \frac{5}{3}\text{div}\{-n_e \varepsilon \mu_e(\varepsilon) E - D_e(\varepsilon)\text{grad}(n_e \varepsilon)\} = -eF_e E - n_e N k_l(\varepsilon) \tag{A4}$$

where $k_l(\varepsilon)$ is the electron energy loss coefficient.

**3 Poisson equation for the electric potential $\varphi$:**

$$\Delta\varphi = -4\pi e(n_p - n_e) \tag{A5}$$

**4 Gas thermal conductivity equation:**

$$\frac{\partial T_{gas}}{\partial t} = \text{div}\{\lambda(T_{gas})\text{grad}(T_{gas})\} + w \tag{A6}$$

where $T_{gas}$ is gas temperature, $\lambda$ the coefficient for thermal conductivity and $w$ is the density of heat sources. For the thermal conductivity coefficient a linear approximation was used:

$$\lambda(T_{gas}) = AT_{gas} + B \tag{A7}$$

It was assumed that there is no convective gas flow in the discharge volume and the local gas density was determined according the ideal gas equation:

$$P = N k_b T_{gas} \tag{A8}$$

The voltage over the discharge was calculated according to:

$$U = U_{emf} - R_b I_d \tag{A9}$$

Where $U_{emf}$ is the voltage of the generator, $R_b$ the external ballast resistance, and $I_d$ is discharge current.

**Boundary conditions:**

*At the cathode z = 0:*

$$F_e = \gamma |F_p|; \quad \frac{\partial F_p}{\partial z} = 0; \quad \varepsilon = \varepsilon_0; \quad \varphi = 0 \tag{A10}$$

*At the anode z = d:*

$$F_e = 1/4(n_e v_{e,th}); \quad n_p = 0; \quad Q_e = 1/4\{n_e v_{e,th}(2k_b T_e)\}; \quad \varphi = U \tag{A11}$$

*And at the lateral walls r = R:*

$$F_e = 1/4(n_e v_{e,th}); \quad \frac{\partial F_p}{\partial r} = 0; \quad Q_e = 1/4\{n_e v_{e,th}(2k_b T_e)\}; \quad \frac{\partial \varphi}{\partial r} = -4\pi\sigma \tag{A12}$$

A symmetry condition was applied at the discharge axis at $r = 0$. Here $\gamma$ is the secondary ion-electron emission coefficient, $\varepsilon_0$ energy of the secondary electrons (assumed to be 1 eV), $v_{e,th}$ the thermal velocity of the electrons, $T_e$ the electron temperature, $Q_e$ the electron energy flux and $\sigma$ the surface charge density.

For the thermal conductivity equation the temperature was assumed to be known (Dirichlet conditions) and equal to 290 K.

A detailed description of the solution procedure for the continuity equations coupled with the Poisson equation in two dimensions can be found elsewhere (see for instance [14, 17] and references therein).

# APPENDIX B

## 1D drift model

This set of equations is analogous to the one used by Ward [11] and can be derived from the continuity condition for the drift current and the electrostatic equation for the electric field. This results in a system of two first order differential equations:

$$\frac{dJ_e}{dz} = \alpha(E) J_e \tag{B1}$$

$$\frac{dE}{dz} = -\frac{4\pi e}{\mu_p E} \{J - J_e(1 + \mu_p/\mu_e)\} \tag{B2}$$

With boundary conditions for the current:

$$J_e(0) = \frac{\gamma \cdot J}{1+\gamma}; \qquad J_e(d) = J \tag{B3}$$

Here $J_e$ and $E$ are the electron current density and electric field respectively; $J$ is the total current density; $\mu_p$, $\mu_e$ are the mobilities of ions and electrons; $\gamma$ is the secondary electron emission coefficient and $\alpha$ is the Townsend's ionization coefficient; $z = 0$ corresponds to the cathode and $z = d$ to the anode location.

For $\alpha$ usually an exponential (Townsend) approximation is used:

$$\alpha/p = C \exp\left\{-D\left(\frac{p}{E}\right)^s\right\} \tag{B4}$$

where the index $s$ is typically equal to 1 or to 1/2.

Following the original work [11] the solution of the system above is found by integrating (B1 and B2) over the discharge gap while varying the assumed electric field value at cathode (or anode) till the boundary conditions (B3) are met. The value of the total current density $J$ is supposed to be known. Standard library routines can be utilized for integration, for instance, in the recent article [3] a reference is given to the ODE-PACK FORTRAN library from the freeware site: netlib.org. It should be noted that the solution of (B1-3) utilizing the described procedure is restricted to low values of current densities. In general the validity of the local model for higher currents in the abnormal branch of the voltage current characteristic is questionable.

**FIGURES:**

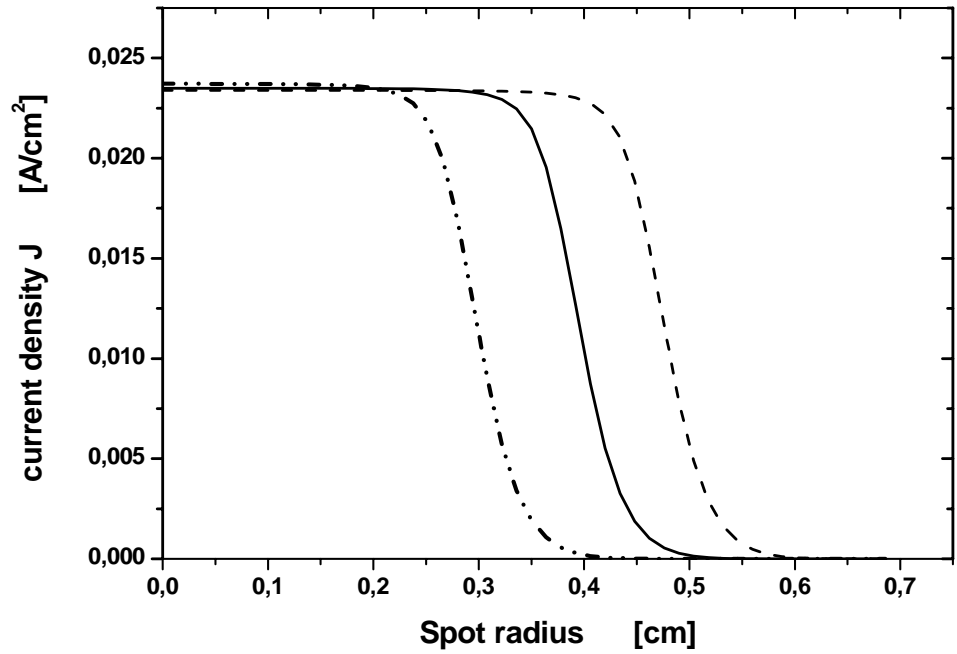

**Fig. 1.** Calculated profiles of the current density vs the spot radius at the cathode for three different values of the total discharge current: $I$ = 7, 12 and 17 mA. (Normal discharge in air at a pressure of 20 Torr and with electrode separation of 1 mm)

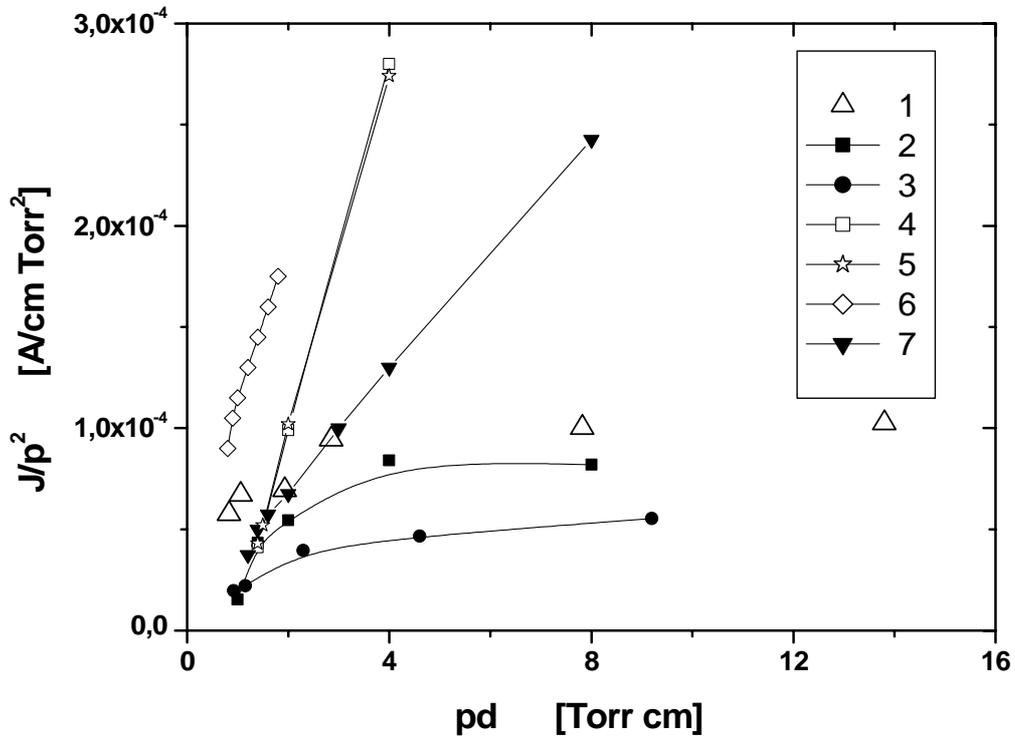

**Fig. 2.** Dependency of the reduced normal current density $J_n/p^2$ on the *pd* parameter for discharge in air: experimental data for 46 Torr (1)[10]; modelling 20 Torr (2); modelling 46 Torr (3); modelling 20 Torr, no gas heating (4), modelling 10 Torr, no gas heating (5), results of 1D drift model with ionization coefficient from Boltzmann code (6), results of 1D drift model, ionization coefficient approximated according [1] (7).

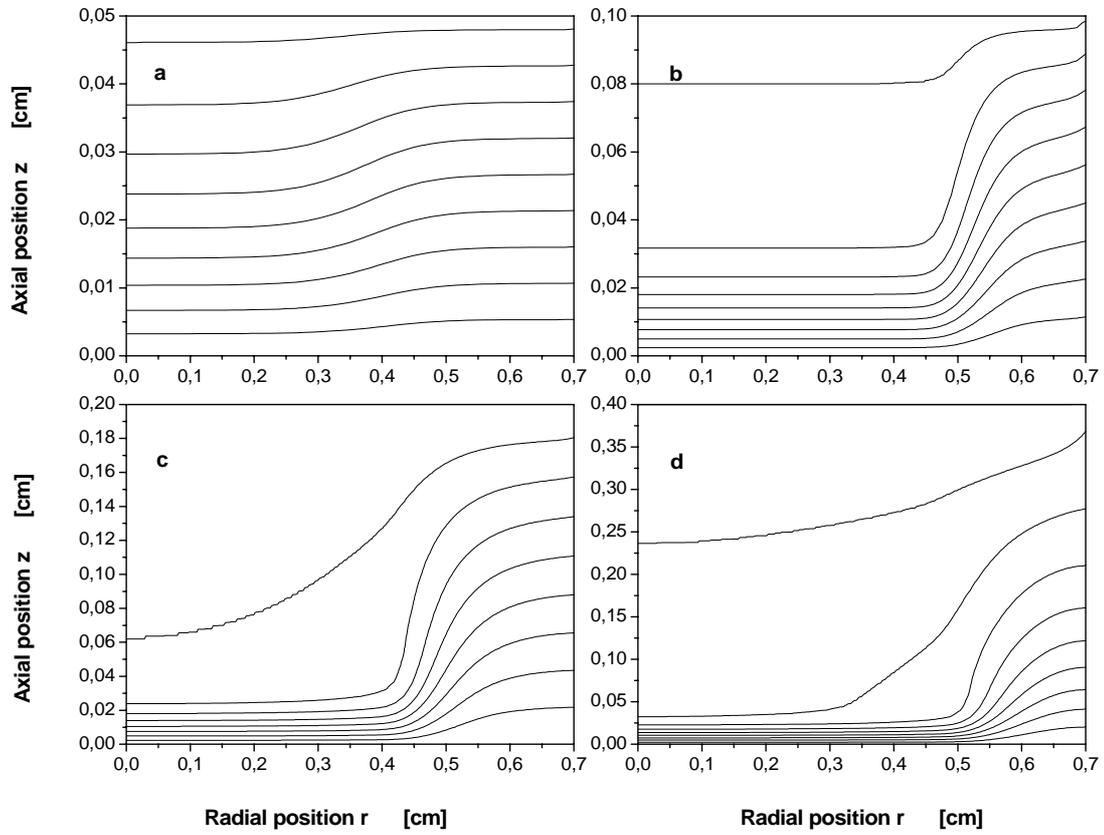

**Fig. 3.** Calculated equipotential contour maps for a normal discharge in air at a pressure 20 Torr. Graphs a, b, c, and d are respectively for electrode separations of 0.5, 1, 2 and 4 mm. z = 0 corresponds to the cathode plane.

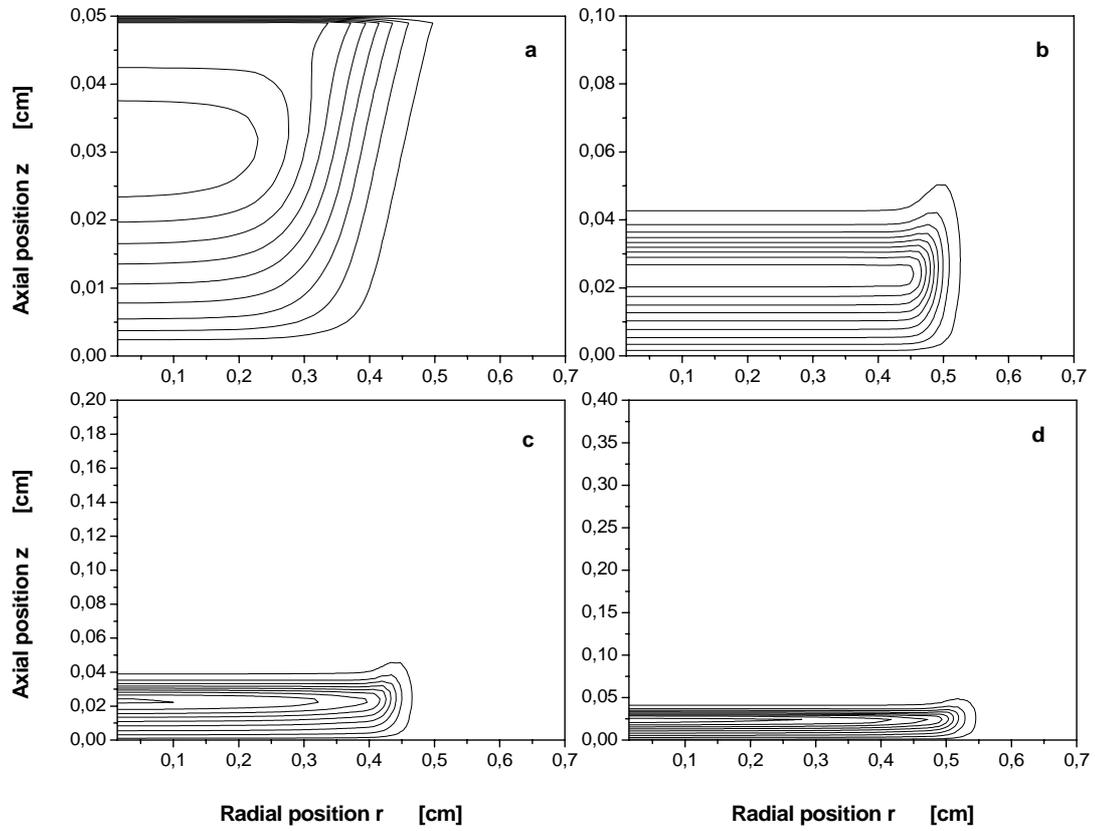

**Fig. 4.** Contours of constant ionization rates ($k_i(\varepsilon)n_e N$ [cm$^{-3}$ / sec]) corresponding to the potential distributions in Fig.3. Graphs a, b, c, and d are respectively for electrode separations of 0.5, 1, 2 and 4 mm. (Normal discharge in air at a pressure of 20 Torr)